\documentclass{kapproc} 

\usepackage{procps} 

\usepackage[dvips]{graphicx}

\upperandlowercase

\kluwerbib 

\begin{document}



\articletitle[]{The Milky Way and the Tully--Fisher relation}


\author{L.~Portinari, J.~Holmberg and C.~Flynn}

\affil{Tuorla Observatory, University of Turku, Vaisalantie 20, 
FI-21500 Piikkio, Finland}

\begin{abstract}
We present an updated estimate of the surface density and surface brightness 
in $B, V, I$ of the local Galactic disc, 
based on a model for the ``Solar cylinder'' calibrated to reproduce 
Hipparcos and Tycho star counts. We discuss the mass--to--light ratio 
of the local stellar disc 
and infer the global luminosity of the Milky Way, which results 
underluminous with respect to the Tully--Fisher relation.
\end{abstract}

\section{Introduction}
\noindent
In the past two decades there has been a steady effort to determine 
the surface mass density of the local Galactic disc, eventually excluding
any significant evidence for disc dark matter (e.g.\ Holmberg \& Flynn 2000, 
2004). As to the local luminosity surface density, 
the infra-red structure of the Milky Way 
seems well understood (Gerhard 2002), but studies in the optical 
typically date back to the '80s (de Vaucouleurs \& Pence 1978; 
Ishida \& Mikami 1982; Bahcall \& Soneira 1980; van der Kruit 1986). 
The surface luminosity and colours
of the ``Solar cylinder'' are useful comparison points in extragalactic 
studies, and to constrain chemo--photometric models of the Milky Way, 
which often serve as a calibration point for the modelling of disc 
galaxies (Boisser \& Prantzos 1999). Estimating the local 
surface brightness across many bands requires accurate knowledge 
of the vertical structure 
of the disc; in this respect, Galactic models have much improved after 
Hipparcos. The time is ripe for an update.

\section{The Galactic model and the Solar cylinder}
\noindent
We derive the surface brightness of the local Galactic disc from the 
stellar census of the Hipparcos and Tycho catalogues, combined
with a model for the vertical structure of the disc.
The model consists of a thin disc, a thick disc 
and a stellar halo (irrelevant in this study). The vertical structure 
is modelled as a series of components (e.g.\ Bahcall 
\& Soneira 1980; van der Kruit 1988)
with a different scaleheight for each component: main sequence
stars of different $M_V$, red giants and supergiants. The $V$ band 
luminosity function and the colour distributions ($B-V$, $V-I$) vs.\ 
$M_V$ are calibrated to reproduce the Hipparcos/Tycho star counts.
The mass model is constrained from the kinematics and the vertical
distribution of AF stars (within 200 pc) and of K giants (both local and out 
to $\sim$1~kpc toward the South Galactic Pole); for details see Holmberg 
et~al.\ (1997); Holmberg \& Flynn (2000, 2004). 

\begin{figure}
\sidebyside{\includegraphics[width=0.4\textwidth,angle=-90]{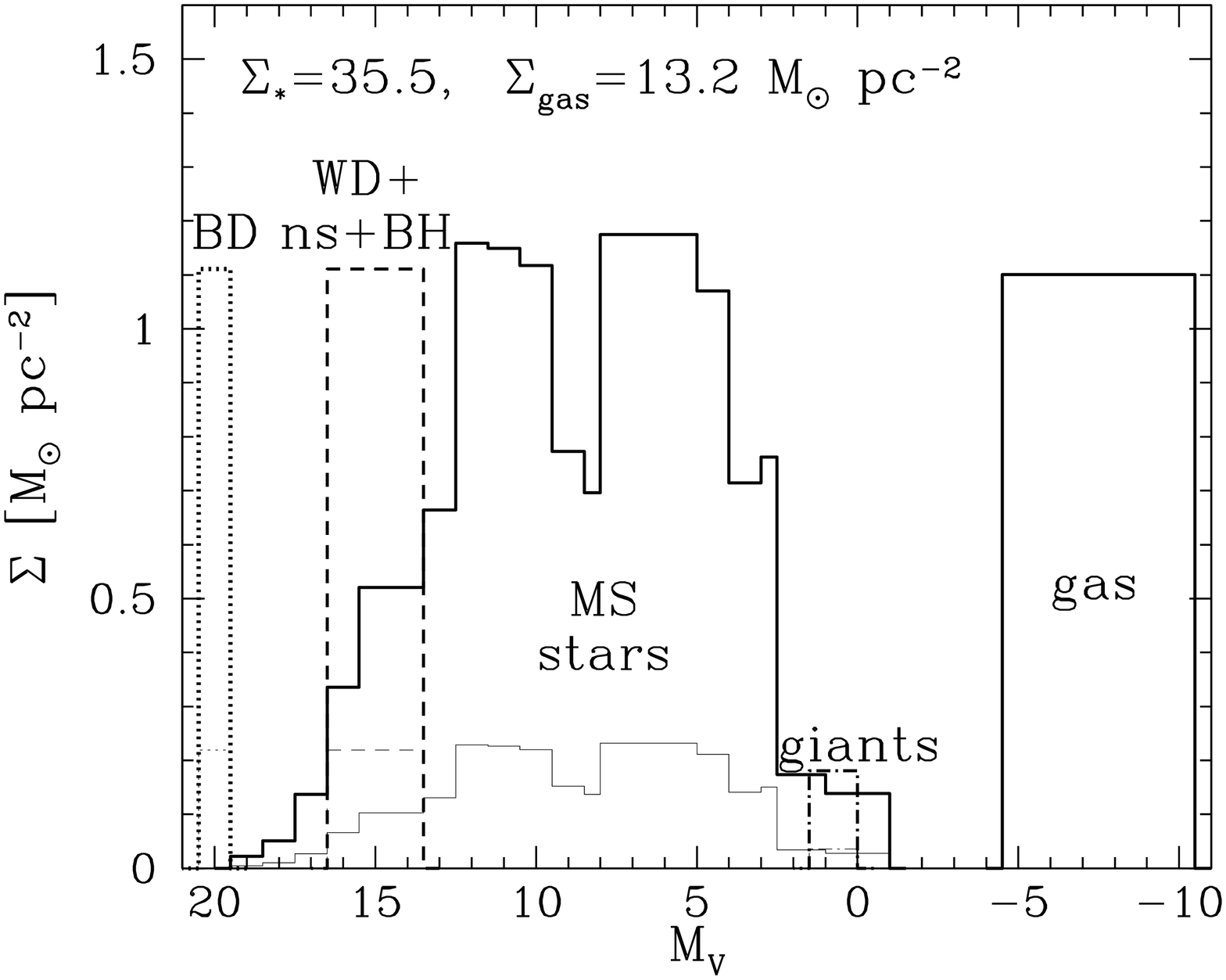}
}{\includegraphics[width=0.4\textwidth,angle=-90]{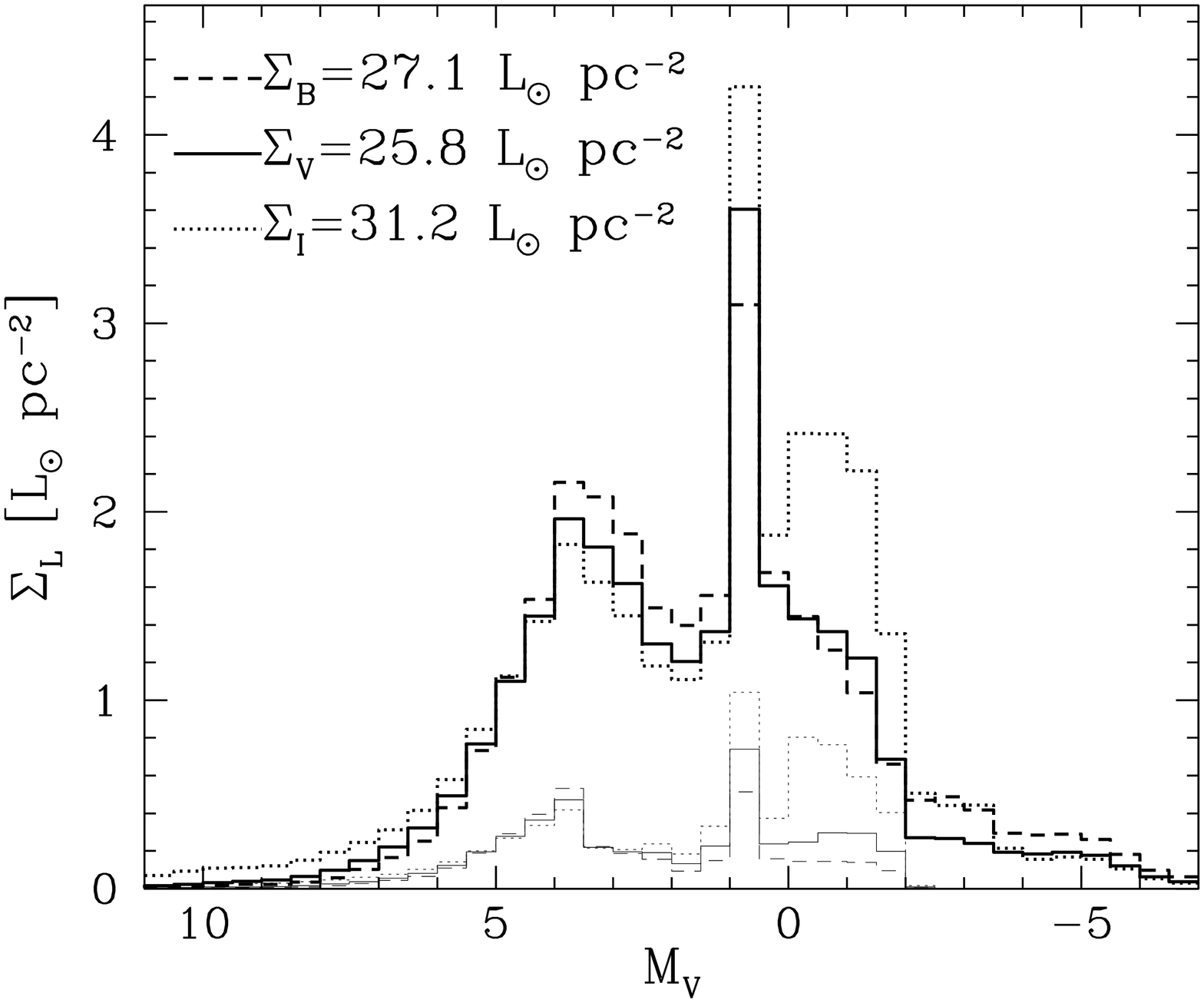}}
\caption{Mass model by Holmberg \& Flynn (2004) and corresponding 
surface luminosity histograms in $B, V, I$. 
{\em Thick lines}: total mass/luminosity; 
{\em thin lines}: thick disc contribution.}
\label{fig:masslummodel}
\end{figure}

Fig.~\ref{fig:masslummodel} shows (a slightly updated version of) 
the mass model of Holmberg \& Flynn (2004), and the corresponding 
$B, V, I$ band luminosity contributions of stars by absolute $M_V$;
the resulting global surface mass/luminosity densities 
are also indicated.
The luminosity peak around $M_V \sim 0.5$ is due 
to clump giants; giant stars contribute 26~(40,~56)~\% of the $B$~($V$,~$I$) 
surface luminosity.

\begin{figure}
\sidebyside{\includegraphics[width=0.55\textwidth]{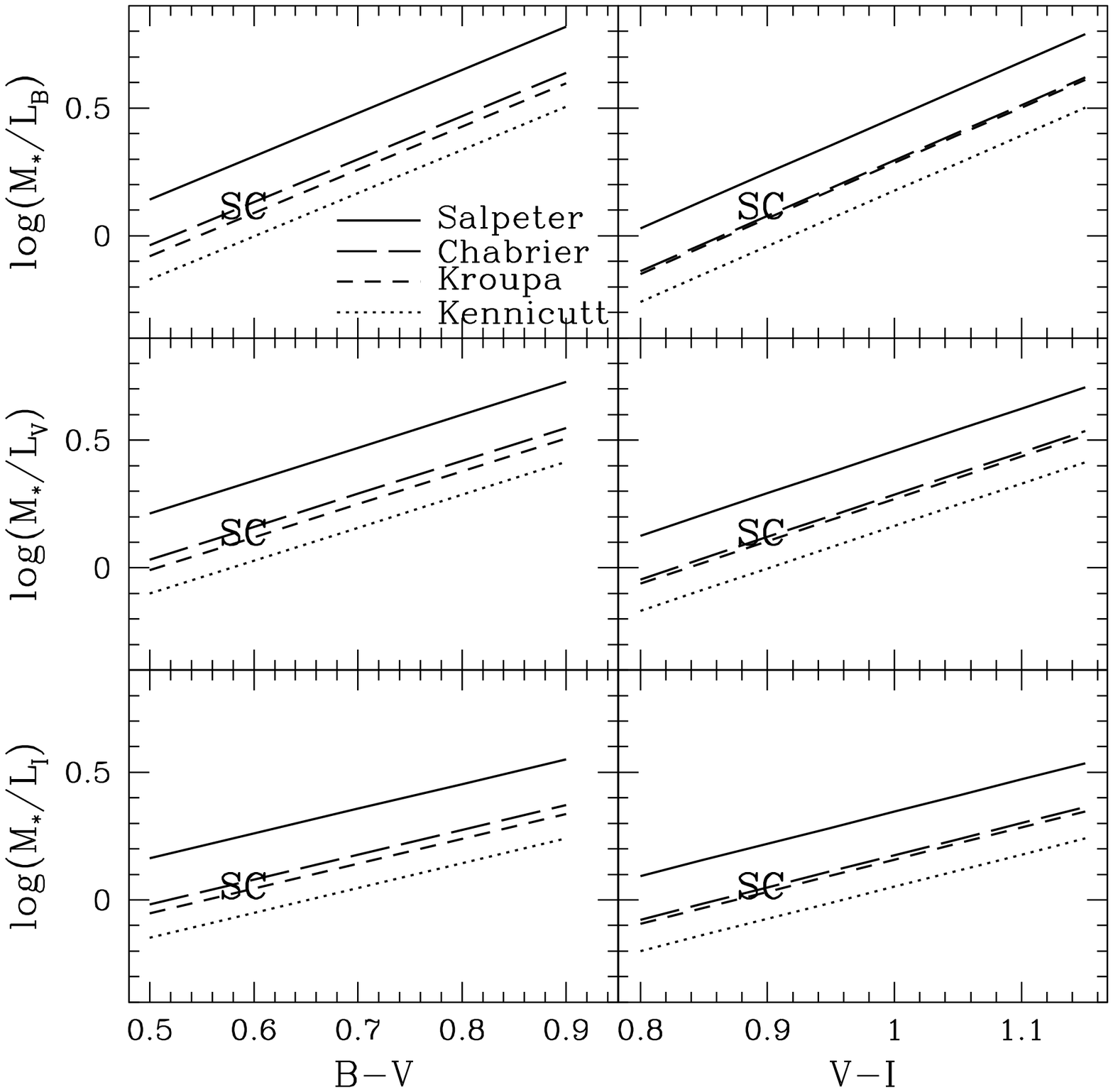}}
{\vspace{-6truecm}
\begin{tabular}{l c c c}
\sphline
 & {\it thin disc} & {\it thick disc} & {\it total} \\
\sphline
$\mu_B$ & 23.7 & 25.7 & 23.5 \\
$\mu_V$ & 23.1 & 24.8 & 22.9 \\
$\mu_I$ & 22.3 & 23.7 & 22.0 \\
$(B-V)$ & 0.53 & 0.90 & 0.59 \\
$(V-I)$ & 0.84 & 1.13 & 0.89 \\
\sphline
\end{tabular}

\bigskip
\begin{tabular}{c c c}
\sphline
 $\frac{M_*}{L_B}$ & $\frac{M_*}{L_V}$ & $\frac{M_*}{L_I}$ \\
\sphline
       1.3         &        1.4        &        1.1        \\
\sphline
\end{tabular}}
\caption{Model results for the surface brightness and colours of the Solar
cylinder, and corresponding location with respect to the colour--M$_*$/L 
relations predicted for different IMFs.} 
\label{fig:colourML}
\end{figure}

Fig.~\ref{fig:colourML} lists the surface brightness and colours 
we derive for the local thin, thick and total disc; and the
stellar mass--to--light ratio (M$_*$/L) of the Solar cylinder.
We also compare our results to the
theoretical colour--M$_*$/L relations predicted by the population synthesis 
models of Portinari et~al.\ (2004) for different Initial Mass Functions (IMFs).
Our SC values are in excellent agreement with the predictions 
for the Kroupa and Chabrier IMFs, which have been derived for
the Solar Neighbourhood --- a successful consistency check.

\section{From the Solar cylinder to the Milky Way}
\begin{figure}
\sidebyside{\includegraphics[width=0.45\textwidth]{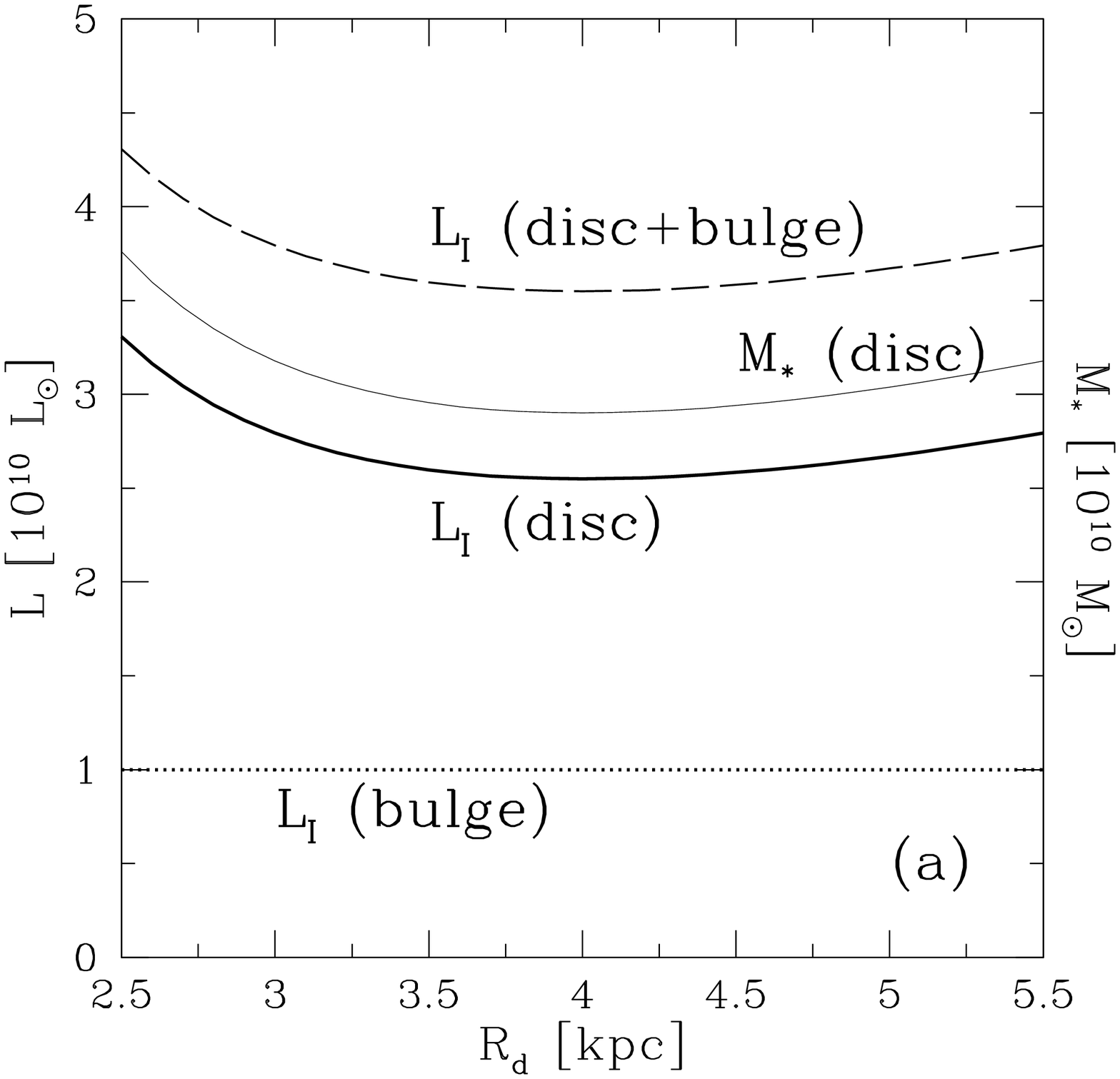}}
{\includegraphics[width=0.45\textwidth]{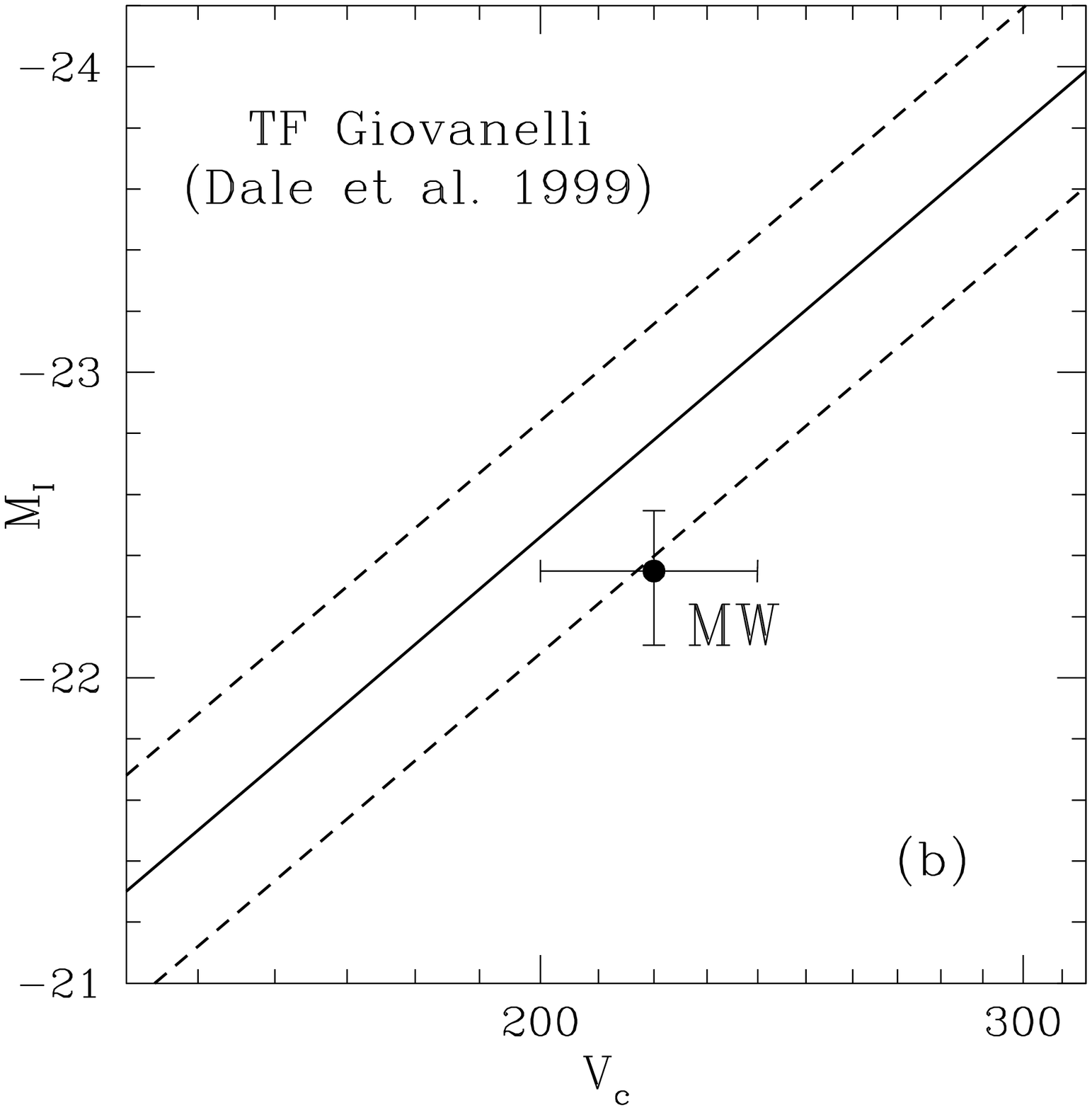}}
\caption{{\em (a)} Total $I$ band luminosity of the Mily Way inferred 
as a function of the assumed disc scalelength $R_d$. {\em (b)} Location
of the Milky Way with respect to the $I$ band TF relation.}
\label{fig:MW-TF}
\end{figure}

\noindent
From the surface luminosity (or density) at the Solar radius one can
infer the total luminosity (or mass) of the Galactic disc
by assuming an exponential radial profile, and the result is quite robust to
the assumed scalelength $R_d$ 
(e.g.\ Sommer--Larsen \& Dolgov 2001). The Solar cylinder probes an inter-arm 
region so we must allow for the spiral arm contrast to derive the actual
azimuthally averaged surface brightness at the Solar radius. 
We will henceforth focus on the $I$ band as this is most typical 
for Tully--Fisher (TF) studies and spiral arms are not expected to induce 
major effects. In $K$ band the spiral arm enhancement is only about 10\%
(Drimmel \& Spergel 2001; Gerhard 2002) and
in the $I$ band the effect should be comparable (Rix \& Zaritsky 1995).
Fig.~\ref{fig:MW-TF}a shows the total $I$ band luminosity 
and stellar mass of the Galactic disc inferred from the local surface
brightness and density, including a 10\% correction for spiral arms. 
We must further add the bulge contribution to get the total luminosity; 
the bulge 
$K$ band luminosity is $\sim 10^{10}$~L$_{\odot}$ (Kent et~al.\ 1991; Gerhard 
2002); we assume the same value in $I$ band, which is probably an 
overestimate as the bulge is mostly composed of older, redder populations. 
The total $I$ band luminosity of the Milky Way is thus 
$\sim 4 \times 10^{10}$~L$_{\odot}$, or
$M_I \sim -22.4$. With a circular speed of $\sim 220 \pm$20~km/sec,
the Milky Way turns out to be underluminous with respect
to the TF relation defined by external spirals (Fig.~\ref{fig:MW-TF}b). 

\section{Conclusions}
\noindent
We have presented an updated estimate of the $B, V, I$ surface brightness 
and colours of the ``Solar cylinder'' 
and reconstructed from the local disc brightness the total luminosity of the 
Milky Way, which appears to be underluminous with respect to the TF
relation; although the discrepancy is not dramatic when considering 
the scatter in the TF relation and the uncertainty in the Milky Way values 
(20\% assumed for the luminosity in Fig.~\ref{fig:MW-TF}). The offset might 
be partly
due to a colour effect, if the Milky Way is a redder and earlier type 
spiral than the Sbc-Sc galaxies defining the TF relation 
(Portinari et~al.\ 2004); however, the disc locally is quite blue 
($B-V \sim 0.6$) which does not argue for significant colour and M$_*$/L 
offsets between the Milky Way and Sbc--Sc spirals.
All in all, the offset might indicate a problem with the luminosity 
zero--point of the TF relation, and/or with the M$_*$/L  
of disc galaxies. A similar conclusion is suggested also by
semi--analytic models of galaxy formation (Dutton et al.\ 2004). 
The issue certainly deserves further investigation.

We are presently refining our $I$ band luminosity estimate with the aid 
of DENIS and other star counts, and defining better our 
errorbars; we are also extending our surface brightness 
determination to other photometric bands, especially in the infrared utilizing
DENIS and 2MASS.







%




\end{document}